\documentclass[11pt]{article}
\usepackage{dsfont}\usepackage{cite}
\usepackage{times} \usepackage{amssymb}
\usepackage{pifont}\usepackage{amsmath}
\usepackage[mathscr]{eucal}
\numberwithin{equation}{section}
\newcommand\remove[1]{}

\newcommand\email[4]{#1{\tiny (at)} #2 {\tiny (dot)} #3 {\tiny (dot)} #4}

\newcommand\ch{\operatorname{ch}}
\newcommand\chz[1]{\operatorname{ch_0}({#1})}
\newcommand\chy[1]{\operatorname{ch_1}({#1})}
\newcommand\chx[1]{\operatorname{ch_2}({#1})}
\newcommand\chw[1]{\operatorname{ch_3}({#1})}
\def\c{\mathcal{C}}

\newcommand\E{{\cal E}}

\def\eq#1{(\ref{#1})}

\def\mF{{\mathscr F}}
\def\I{{\cal I}}
\newcommand\II{\mathds{1}}
\def\J{{\mathscr J}}
\def\goth#1{{\mathfrak #1}}
\newcommand\h{\goth{h}}
\renewcommand\H{\goth{H}}
\renewcommand\v{\goth{v}}

\newcommand\ie{{\slshape i.e.~}}
\newcommand\mint{\goth{m}}
\newcommand\nint{\goth{n}}
\def\N{\mathcal{N}}
\def\T{{\mathds{T}}}
\def\viz{{\slshape viz.~}}
\def\Z{\mathds{Z}}
\begin{document}
\title{Moduli stabilization with non-Abelian fluxes }
\author{ 
Alok Kumar, 
\thanks{\email{kumar}{iopb}{res}{in}} 
\hfil Subir Mukhopadhyay, 
\thanks{\email{subir}{iopb}{res}{in}} \\ 
\small Institute of Physics, Bhubaneswar 751~005, India.
\and 
Koushik Ray \thanks{\email{koushik}{iacs}{res}{in}} \\
\small Department of Theoretical Physics,\\
\small  Indian Association for the Cultivation of Science.
\small  Calcutta 700~032. India.}
\date{}
\maketitle
\vfil
\begin{abstract}
\noindent 
We study stabilization of moduli in the type--IIB 
superstring theory on the six-dimensional 
toroidal orientifold  $\T^6/\Omega\cdot(-1)^{F_L}\cdot\Z_2$. 
We consider background space-filling D9-branes wrapped on the orientifold
along with non-Abelian fluxes on its world-volume 
and demonstrate with two examples that
this can stabilize all the complex structure moduli
and some of the K\"ahler moduli. 
\end{abstract}
\thispagestyle{empty}
\clearpage
\section{Introduction}
Superstring theories live in ten space-time dimensions. 
Compactifying on a six-dimensional space yields a theory 
in four dimensions, which are
identified with the four dimensions of the real world.
Any such consistent compactification yields a vacuum of string theory
and there exists plethora of such vacua which have been studied in a 
variety of contexts over the last few decades. 
String compactification generically leads to a large number of flat moduli,
which should have fixed values in a realistic model. 
Fixing the values of moduli to obtain a stable vacuum
received the appellation \emph{moduli stabilization}. 
Certain progress in understanding various mechanisms 
of moduli stabilization in different string theories has 
been achieved in the past few years. Among others, there 
are now schemes for moduli stabilization using background fluxes 
\cite{Gukov:1999ya,Giddings:2001yu} as well as background D-branes 
with fluxes
\cite{Bachas:1995ik,Angelantonj:2000hi,Blumenhagen:2003vr,Cascales:2003zp}, 
using intersecting brane 
configurations\cite{Blumenhagen:2005mu} or branes at angles\cite{Cvetic:2001nr}, construction of stable 
bundles in Heterotic string theories, to mention a few. Moreover, 
some of these are
related through the duality symmetries of string theory.
While the question of whether nature has chosen one from 
\emph{these} nimiety of
vacua for us to live in and if so, then whether it was a 
`principled' choice 
or a capricious one still awaits a decisive answer, 
the various constructions of moduli stabilization have 
interesting features which attracted special attention.

The most explored scheme of moduli stabilization uses 
background fluxes leading to $\N=1$ supersymmetric
vacua, known as flux vacua.
In a flux vacuum the NS-NS and RR fluxes in 
the internal space, through the
consistency requirements of string theory, 
can fix the values of the moduli. In particular,  in the 
type--IIB closed string theory this mechanism can stabilize 
all the complex structure
moduli and the dilaton-axion modulus to fixed values 
\cite{Kachru:2002he, Gorlich:2004qm, Kachru:2003aw}.
In certain examples \cite{Kachru:2003aw} it has
even been possible to obtain meta-stable vacua akin to 
the de-Sitter space-time, deemed to be a rather realistic 
space-time of late. These fluxes, however, do not render
themselves to an exact world-sheet description.
Moreover, in the type--IIB theory
fixing the K\"ahler moduli using such
fluxes calls for non-perturbative means, restricting the  
viability of such analyses in the
effective supergravity theory
\cite{Gorlich:2004qm, Kachru:2003aw}.

Recently, another mechanism of moduli stabilization 
has been proposed employing constant Abelian magnetic 
fluxes on the world-volume of a background D-brane 
wrapped on a six-torus, $\T^6$, or its orbifolds 
\cite{Antoniadis:2005nu, Antoniadis:2004pp, 
Bianchi:2005yz, Bianchi:2005sa}
( for the compactification of type--I string theory
on smooth Calabi-Yau with non-Abelian bundle
from a slightly
different perspective see \cite{Blumenhagen:2005pm,
Blumenhagen:2005zg,Blumenhagen:2005zh}). 
The analysis of the Abelian fluxes is exact 
in the open string theory 
and thus not restricted to the lowest order in the inverse 
of the string tension $\alpha'$.
These constant magnetic fluxes stabilize many of the 
complex structure as well as K\"ahler
moduli 
\cite{Antoniadis:2005nu, Antoniadis:2004pp}. 
In this article we study further examples 
in this class of schemes. 
We generalize the construction 
of magnetic branes on toroidal orbifold
by turning on
constant non-Abelian magnetic fluxes on the 
two-cycles of the internal space, as
opposed to the Abelian fluxes.

The analysis of magnetic branes with non-Abelian fluxes 
is, again, plagued with the vice of not rendering itself to an exact 
string theoretic derivation 
\cite{Bianchi:2005yz, Bianchi:2005sa}.
However, in stabilizing the the totality of available moduli, 
the open string moduli are to
be coupled with the closed string moduli, 
with the latter being treated at but the lowest
order in any known scheme. 
Hence, in want of a better method of stabilization, 
which treats both sides exactly, the present analysis, though limited,
is quite relevant.

The paper is structured as follows. In the next section 
we briefly discuss the salient features  
of moduli stabilization with magnetic branes. 
In section \ref{secmodel} we illustrate the 
stabilization of complex structure moduli and K\"ahler moduli. 
We conclude with a discussion in the section \ref{secdis}.
Some of the notations and conventions are elaborated in Appendix \ref{appndx}.
\section{The constraints}
Let us begin with a brief discussion of the conditions of supersymmetry
preservation for D-branes. 
Let us consider the type--IIB theory compactified on 
a six dimensional variety $X$. D-brane configurations are
supersymmetric if they are wrapped on 
supersymmetric cycles of $X$. In the presence of magnetic fluxes in
the world-volume of the D-brane this 
condition gets further modified.
For a single space-filling D-brane 
with magnetic flux one can write
a $\kappa$-symmetric action 
\cite{Bergshoeff:1997kr,Antoniadis:2005nu,
Antoniadis:2004pp} 
from which follows the BPS condition
\cite{Bergshoeff:1997kr
} 
\begin{equation}\label{susy0.1}
(1-\Gamma)\eta = 0,
\end{equation}
for the fermion $\eta$, with
$\Gamma$ given as
\begin{equation}
\Gamma = \frac{\sqrt{|g|}}{\sqrt{|g+F|}}
\sum_{n=0}^\infty \frac{1}{2^n n!} 
\;\gamma^{\mu_1\nu_1 \cdots \mu_n\nu_n}
F_{\mu_1\nu_1\cdots \mu_n\nu_n}J^{(n)}_9, \label{susy0.2}
\end{equation}
where $g$ denotes the metric induced on the world-volume of a D9-brane, 
$F$  the Abelian field strength on the world-volume and $|\cdot|$ denotes a
determinant. Moreover, 
\begin{equation}
J^{(n)}_9 = (-1)^n\sigma_3^{n+3} i\sigma_2\otimes\gamma^{(11)}, \qquad
\Gamma^{(11)} = \frac{i}{10!\sqrt{|g|}}
\epsilon^{\mu_1\dots\mu_{10}}\gamma_{\mu_1\dots\mu_{10}},
\end{equation}
where $\sigma$'s are the Pauli matrices, $\gamma$'s 
are the ten-dimensional $\gamma$-matrices 
and $\epsilon^{\mu_1\dots\mu_{10}}$ denotes the ten-dimensional 
antisymmetric tensor indicating the choice of the orientation
of $X$. This expression contains the perturbative terms
of all orders in $\alpha'$ but liable to receive
corrections from world-sheet instantons. 

In order to obtain a supersymmetric theory 
upon orientifolding, the supersymmetry
preserved by the D-brane 
configuration needs to be the same as
the one preserved by the orientifold plane.
It follows from the Dirac-Born-Infeld action
\cite{Marino:1999af,Antoniadis:2004pp,Antoniadis:2005nu}
that supersymmetry is preserved if
the following two conditions are met,
\begin{gather}
\mF_{ij}=0 \label{susy1.1}, \\
\mathrm{Im}\big(e^{i\theta}\!\int e^{(\mF^{(1,1)}-i\J)^3}\big) = 0 
\label{stability0},
\end{gather}
where $\mF^{(2,0)} = \mF_{ij}$ and $\mF^{(1,1)}=\mF_{\bar\imath\bar\jmath}$ denote the
$(2,0)$-forms and $(1,1)$-forms, respectively, representing the
constant Abelian magnetic fluxes, while $\J$ denotes the K\"ahler form 
and a product of forms is to be interpreted in terms
of wedge products.
The value of the parameter $\theta$ depends on the kind of orientifolding
performed. For O3- and O5-planes, $\theta=0$ and $\theta=\pi$, respectively.
Since the two equations (\ref{stability0}) and
(\ref{susy1.1}) depend on the
complex structure moduli and K\"ahler moduli, 
respectively, they stabilize
the respective moduli, as has been demonstrated in models with constant Abelian
magnetic fluxes\cite{Antoniadis:2004pp,Antoniadis:2005nu}.

We consider a generalization of this mechanism
to cases with constant non-Abelian magnetic fluxes.
The supersymmetry preserved by D-branes with non-Abelian 
magnetic fluxes on their world-volume has been 
discussed earlier \cite{Douglas:2001ug} in the context of  
$\N=2$ theories, where the internal space $X$ is Calabi-Yau. 
In this article we consider
$X=\T^6/\Omega\cdot(-1)^{F_L}\cdot\Z_2$. For the case at hand
we need three ingredients from the various facets of the conditions of
preservation of supersymmetry. The first ingredient is 
the requirement that the vector bundle
$\E$ on the world-volume of the D-brane, that describes the D-branes
with non-Abelian fluxes, is holomorphic.
The condition on the fluxes is \eq{susy1.1}, now with a 
non-Abelian $\mF^{(2,0)}$. 

The second ingredient is the requirement of equality of the phases 
of supersymmetries
preserved by the D-branes and the orientifold plane.
The phase of the supersymmetry preserved by a D-brane
is given by the grade $\phi(\E)$ of $\E$, which is related to
the central charge $Z[\E]$ of the associated D-brane configuration as
\begin{equation}
\phi(\E) = \frac{1}{\pi} \text{Im}{(\log Z[\E])}.
\label{grade}\end{equation}
Thus, the condition of equality of the phases assumes the form,
\begin{equation}
\text{Im}{(e^{-i\theta} Z[\E])} = 0 \label{stability1},
\end{equation}
which generalizes \eq{stability0}.
In order to see this let us recall that 
\cite{Douglas:2001ug} the central charge of  D-branes 
wrapped on a six-dimensional compact space $X$ is determined by
their RR charges, which can be obtained from a Wess-Zumino term in the
world-volume action $\int C_{(10-2i)}\wedge \ch_i{\E}$, where $\ch_i(\E)$ denotes the
$i$-th Chern character of the bundle $\E$ and $C_i$ denotes an RR $i$-form coupled to
it.  In the large volume limit the 
central charge for pure D-$2p$-branes becomes the volume, given by
\cite{Harvey:1996gc,Minasian:1997mm,
Freed:1999vc,Aspinwall:2001dz},
\begin{equation}\label{centralcharge}
Z[\E] =\int\limits_X e^{-i\J}\wedge\ch({\E}).
\end{equation}
For a single brane $\E$ is a line bundle and the central charge reduces to
$Z[\E] = \int\limits_X (\mF^{(1,1)}-i\J)^3$ and hence \eq{stability1} 
reduces to \eq{stability0}. Using S-duality, the perturbative part of this condition of supersymmetry preservation, namely, that of \eq{stability0} 
has been shown\cite{Blumenhagen:2005pm} to be related to vanishing 
of FI terms in compactification of SO(32) Heterotic
string theory.

For the case at hand, therefore, the
supersymmetry preserving D-brane satisfies \eq{susy1.1} with non-Abelian fluxes
and 
\begin{equation}
\text{Im}{\big(e^{i\theta}\!\int\limits_X e^{-i\J}\wedge\ch{\E}\big)} = 0 \label{susy1.2},
\end{equation}
obtained from \eq{centralcharge}
which can be expanded to
\begin{equation}
\tan{\theta}
\big( \J\wedge\J\wedge\chy{\E} - \chw{\E}
\big) = 
\big(
\J\wedge\J\wedge\J\chz{\E} - \J\wedge\chx{\E}
\big). \label{susy1.3}
\end{equation}
Since we shall consider an O3-plane, \ie $\theta=0$, 
equation \eq{susy1.2} reduces to
\begin{gather}
\J\wedge\J\wedge\J \chz{\E} = \J\wedge\chx{\E}. \label{jjj}
\end{gather}
In addition, as we need $\theta$ to be the correct phase of
supersymmetry, we are also required to abide by
\begin{gather}
\J\wedge\J\wedge\chy{\E} < \chw{\E}. \label{fff}
\end{gather}

There is an additional 
condition for the preservation of supersymmetry of the 
D-brane configuration, which is the third ingredient. 
It requires that the bundle $\E$ is $\Pi$-stable.
That is, the D-brane configuration described by the vector
bundle $\E$ is stable with respect to the decay
$\E \rightarrow \E_1 \oplus \E_2$ if 
$\phi(\E_i)<\phi(\E)$ for $i=1,2$,
where $\phi(\E)$ is the grade defined in \eq{grade}. 
When $\phi(\E) =\phi(\E_1)=\phi(\E_2)$,
the D-brane configuration is said to be marginally 
stable with respect to this decay. 
In the limit of vanishing string length, $l_s\rightarrow 0$, 
the $\Pi$-stability condition
reduces to the condition for the existence of 
solutions to the Hermitian-Yang-Mills equations
\cite{Harvey:1996gc,Douglas:2001ug}. 
In this article we shall refrain from discussing this condition in 
any further detail, save for a few comments on satisfying it 
in the final section.

There are further conditions, in addition to the ones arising 
from supersymmetry preservation, hitherto discussed. 
These originate from the requirement of tadpole cancellation,
that is, the vanishing of the total RR charge of each of the RR fields
on the compact manifold.
The coupling of a D9-brane to the various RR fields can be obtained
from the Wess-Zumino part of the world-volume action, mentioned above, namely,
\begin{equation}
S_{WZ} = \int C_4\wedge\chw{\E} +
C_8\wedge\chy{\E}.
\end{equation}
Thus, in the presence of fluxes corresponding to bundles 
with non-vanishing first and third Chern
characters, a D9-brane acquires charges of D7- and D3-branes respectively. 
Two more restrictions are to be imposed to prohibit this.
First, we require that the charge equivalent to that of a D7-brane, as a D9-brane wraps a
two-cycle ${\cal C}_2$ of $X$ vanishes, that is,
\begin{equation}  \label{d7tadpole}
\int\limits_{{\cal C}_2} \chy{\E} = 0, 
\end{equation} 
for all two-cycles of $X$.
Similarly, requiring that the equivalent of the charge of a D3-brane vanishes for the
brane-orientifold configuration means that the total D3-brane charge on the D9-brane,
as the latter wraps X, cancels the charges of the sixty four O3-planes positioned at
the sixty four fixed points of $\T^6/\Z_2$, each contributing the equivalent of
a quarter of a D3-brane
charge with an opposite sign, that is,
\begin{equation} 
\int\limits_X \chw{\E} = 16.\label{d3tadpole}
\end{equation} 

To summarize, we impose the conditions \eq{susy1.1}, \eq{jjj}, \eq{fff}, \eq{d7tadpole}
and \eq{d3tadpole} on the bundle $\E$. 
Over and above, there are certain K-theoretic constraints
\cite{Maiden:2006qe} to be satisfied for building a consistent model. 
We shall not delve into a discussion of these conditions,  
which require the vanishing  of the second 
Chern characters modulo an integer. 
To keep our discussion simple we shall choose the branes to wrap a cycle only once.
The K-theoretic constraints can
be taken care of by choosing non-zero wrapping numbers.

Even with this simplicity, we show that it is possible to 
construct consistent models that has  
all the complex structure moduli and some of the K\"ahler moduli
stabilized. The axion-dilaton moduli remains unfixed. 
It may be possible to stabilize it by turning on
NS-NS and R-R fluxes \cite{Kachru:2002he}. 
While these may be generalized, they are not marred by 
considering the K-theoretic constraints. 

Let us close this section by mentioning the  
quantization conditions 
on the fluxes relevant to the D9-brane configuration
\cite{Bianchi:2005sa,Bianchi:2005yz,Bianchi:1991eu 
,Bianchi:1997rf,Witten:1997bs,
Kakushadze:1998bw ,Angelantonj:1999jh}.
Let us consider the embedding of the brane into the compact space $X$
given by
\begin{equation}
z^i = {e}^i_\alpha \sigma^\alpha,
\label{gothw}
\end{equation}
where a $\sigma$ denotes a world-volume co-ordinate of the brane.
The matrix $e$ encodes the information of the number of times a brane wraps a cycle of
$X$. If the branes wrap the cycles more than once, one has to count the cycles with
(integral) multiplicities in evaluating the pairings $\int\ch(\E)$ on the cycles.
Demanding, then, that the Euler character of $X$ is integral, corresponding to the
quantization of fluxes, lets us choose the Chern
characters to assume values in the rational cohomology, rather than the integral one. 
However, for the sake of simplicity, we  have chosen 
${e}^i_\alpha = \delta^i_\alpha$, corresponding to unit wrapping number of branes and
hence work with integral cohomology.
\section{The models}\label{secmodel}
In this section we consider moduli stabilization
in an orientifold of the type--IIB theory compactified on a six-torus, $\T^6$.
The orientifolding action is given by $\Omega\cdot\Z_2\cdot(-1)^{F_L}$ 
where $\Omega$ denotes the world-sheet parity reversal operator, 
$\Z_2$ flips signs of all the co-ordinates of $\T^6$ and $(-1)^{F_L}$ 
changes the sign of the Ramond vacuum on the left.

Let us start with a general gauge-theoretic description
of the configuration. In the models we discuss below, we consider stacks of D-branes,
the number of stacks are two in both the models, but can be more in general
\cite{Antoniadis:2004pp,Antoniadis:2005nu}.
A stack corresponds to a direct summand of the bundle $\E$. Generally, 
we consider $N$ number of D9-branes
wrapped on $X$ in each stack, where $N$ may differ from one stack to another. 
The configuration corresponding to one stack is described by a $U(N)$
gauge theory, corresponding to a bundle of rank $N$
along with a magnetic flux.
Denoting the field strength of the flux by $\mF$ in the complex basis of $z$'s
and by $F$ in the real basis of $x, y$'s, as given in Appendix \ref{appndx},  
we can write the components of the field strength in the real 
basis as
\begin{alignat}{6}
F_{x_ix_j} &= f_{ij} &\quad F_{x_iy_j} &= g_{ij} &\quad F_{y_iy_j} &= h_{ij} \notag\\
\label{fgh}
&= f^a_{ij}T^a, &\quad
&= g^a_{ij}T^a, &\quad
&= h^a_{ij}T^a,
\end{alignat}
where $T^a$, $a=0,1,\dots, n^2-1$, denote the generators of $U(N)$. 
The matrices $f^a$ and $h^a$ are anti-symmetric.
In this notation the field strengths
$\mF^{(2,0)}$ and $\mF^{(1,1)}$ become
\begin{gather}
\mF_{ij}
= [(\I - \bar{\I})^{-1})]^{\text T}
\big(
{\bar{\I}}^{\text T}f\bar{\I} -
{\bar{\I}}^{\text T}g +
g^{\mathrm{T}}~{\bar{\I}} + h
\big)
(\I - \bar{\I})^{-1}, \label{F20} \\
\mF_{i\bar{\jmath}}
= - [(\I - \bar{\I})^{-1}]^{\text T}
({\bar{\I}}^{\text T}f\I -
{\bar{\I}}^{\text T}g +
g^{\text T}\I + h) 
(\I - \bar{\I})^{-1} \label{F11},
\end{gather}
where $\I$ denotes the complex structure matrix and a bar designates complex
conjugation.
Before constructing a flux configuration
with the constraints discussed in the previous section
let us impose the simpler restrictions needed
to be satisfied.
First, we want to stabilize the complex structure and fix it to $\I =i\cdot\II_3$,
where $\II_n$ denotes the $n\times n$ identity matrix.
In order to obtain the flux configuration that stabilizes the
complex structure to this value let us substitute  
$\I=i\cdot\II_3$ in the expression for 
$\mF^{(2,0)}$ given in \eq{F20} and set that equal 
to zero, thereby guaranteeing the holomorphicity of the two-forms.
The restrictions ensuing from this condition are
\begin{equation} 
f_{ij}=h_{ij} \qquad g_{ij}=g_{ji}.
\end{equation} 
Using these and setting 
the complex structure to $\I = i\cdot\II_3$ in \eq{F11} simplifies the 
expression of $\mF^{(1,1)}$ to
\begin{equation}
\mF_{i\bar{\jmath}} = \frac{1}{2} (f_{ij}+ig_{ij}).
\end{equation}
In terms of the basis of $(1,1)$-forms chosen in Appendix \ref{appndx}
it becomes 
$\mF_{i\bar\jmath} = [g_{ij}\h^{+}_{ij} + 
f_{ij}\h^{-}_{ij}]$.
In order to avoid unnecessary complications we choose to fix the 
K\"ahler $(1,1)$-form to be a simple one, namely,
$\J = (i/2)J\delta_{ij}dz^i\wedge d{\bar z}^{\bar\jmath} 
= J\h_{kk}^{+}$. We shall discuss more about the stabilization of $\J$ to this
value shortly.
Let us now present two models and discuss the further
issues of moduli stabilization within the
context of these specific models only.
\subsection{Model I}
In this first model we consider two stacks. Although consistent, from 
stabilization point of view, this model turns out to be of limited use due
to the large 3-brane 
tadpole contribution from just a single pair of stacks. However, 
this discussion helps set up our notations which we use for more 
practical use in $\S\S$~\ref{modelII}.
The stacks, represented by two bundles $\E_1$ and $\E_2$, are both $U(2)$ 
stacks with only an Abelian flux turned on in the former while the latter has
a non-Abelian part as well. The fluxes through the various cycles of $X$ are
specified completely by specifying the matrices $f$, $g$ and $h$. We 
shall often suppress the
stack indices in these matrices, when the stack they pertain to is 
clear from the
context. The fluxes in the first stack are
\begin{equation}
g_{ij} = \lambda\delta_{ij}\II_2,\qquad
f_{ij} = f\epsilon_{ijk}\eta_k\II_2,\qquad
|\eta|^2 = \sum_{i=1}^3 \eta_i^2,
\end{equation}
where $\lambda$, $f$, and
$\eta_i$, $i=1,2,3$, are some constant parameters. The moduli are fixed in terms of these
parameters.
In the basis of $(1,1)$-forms chosen in Appendix \ref{appndx}, 
$\mF^{(1,1)}$ can be written as
\begin{equation}
\mF = [f\epsilon_{klm}\eta^m \h_{kl}^{-}
+ \lambda\h_{kk}^{+}]\II_2.
\end{equation}
The Chern characters of $\E_1$ for such a choice of  magnetic fluxes are
\begin{equation} 
\begin{split}
\chz{\E_1} &= 2,\\
\chy{\E_1} &= 2[\lambda\h^{+}_{kk}
+ f \epsilon_{klm}\eta_m \h^{-}_{kl}],  \\
\chx{\E_1} &=  2 \big[f^2 \eta_k\eta_l\H^{-}_{kl}
+ f\lambda\epsilon_{klm}\eta_m
\H^{+}_{kl} -\lambda^2\H^{-}_{kk}\big], \\
\chw{\E_1} &= 2 \lambda 
\big( \lambda^2 - f^2 |\eta|^2 \big) \v.
\label{chI}
\end{split}
\end{equation} 
Substituting the expression  
$\J =J\h_{kk}^{+}$ of the K\"ahler two-form
and the above Chern characters in \eq{jjj} and \eq{fff} we
obtain the following relations among the parameters, 
\begin{gather}
J ( 3\lambda^2 - f^2 |\eta|^2 ) =  J^3, 
\label{jjj1}\\
2 \lambda ( \lambda^2 - f^2|\eta|^2 - 3 J^2 ) > 0.
\label{fff1}
\end{gather}
The second stack, represented by a bundle $\E_2$, is also with a 
$U(2)$ gauge group. This time, however,
we choose the flux to possess an Abelian as well 
as a non-Abelian $SU(2)$ part.
The fluxes in this stack are, 
\begin{gather}\label{e311}
g_{ij} = -\lambda\delta_{ij}\II_2 + g\frac{\xi^a}{2}\delta_{ij}T^a,\quad
f_{ij} = -f\epsilon_{ijk}\eta_k\II_2 + g\epsilon^a_{ij}T^a,\quad
|\xi|^2\equiv\sum_{a=1}^3 (\xi^a)^2 = 4\end{gather}
where $T^a, a=1,2,3$, denote the three generators of the $SU(2)$ group
chosen to satisfy $\text{tr}(T^a T^b)=2\delta^{ab}$, so that these can be
taken to be the Pauli matrices. We have also introduced new parameters $g$ and
$\xi^a$, $a=1,2,3$, for this stack.
We have chosen the Abelian part in this second stack to be
equal and opposite to that in the first stack, thus ensuring the vanishing of the
total D7-brane charge.
The restriction given by the third equation in \eq{e311}
on the parameters $\xi^a$
is necessitated by the requirement that both the
stacks satisfy \eq{jjj} for the same value of the
K\"ahler modulus $\J$.
The magnetic fluxes for this stack can be written in terms of the
basis $(1,1)$-cycles as
\begin{equation}
\mF^{(1,1)} = - [
\lambda\h_{kk}^{+} + f\epsilon_{klm}\eta_m\h_{kl}^{-}]\II_2
+ g[\frac{\xi^a}{2}\h_{kk}^{+} + \epsilon_{akl}\h_{kl}^{-}]T^a.
\end{equation}
The various Chern characters for the second
stack are given by
\begin{equation} 
\label{chII}
\begin{split}
\chz{\E_2} &= \chz{\E_1},\\
\chy{\E_2} &=-\chy{\E_1}, \\
\chx{\E_2} &= \chx{\E_1} +g^2 \epsilon_{ijk}\xi^k\H^{+}_{ij},\\
\chw{\E_2} &= -\chw{\E_1} + 2g^2f (\xi\cdot\eta)\v,
\end{split}
\end{equation} 
where $\xi\cdot\eta= \sum\limits_{i=1}^3\xi^i\eta_i$. 
Substituting these Chern characters in \eq{jjj} 
leads to \eq{jjj1}, the same condition as  
was obtained for the first stack. However, \eq{fff} now leads to a new restriction on
the parameters, namely,
\begin{gather}
\label{fff2}
g^2f(\xi\cdot\eta) - \lambda(\lambda^2 - f^2|\eta|^2 - 3J^2 ) > 0.
\end{gather}
Comparing \eq{fff1} and \eq{fff2} we see that if we
turn off the non-Abelian part of the second stack by setting
$g=0$, then these two equations can not be simultaneously satisfied.
This has been one of the
reasons for introducing the non-Abelian flux in the second stack.

Now combining the equations \eq{jjj1},\eq{fff1} and \eq{fff2}) we derive 
\begin{equation}
J^2 = 3\lambda^2 - f^2|\eta|^2,\quad
g^2f (\xi\cdot\eta) > \lambda(\lambda^2 - f^2|\eta|^2 - 3 J^2) > 0.
\label{fff03}\end{equation}
We still have the tadpole cancellation conditions \eq{d7tadpole} and
\eq{d3tadpole} to impose. The contribution of the pair of stacks
to the D3-brane tadpole is,
\begin{equation}
\rho_2 = \chw{\E_1} + \chw{\E_2} = 2g^2f(\xi\cdot\eta).
\label{d33}\end{equation}
Combined with \eq{fff03} and \eq{d33} this
imposes a condition on the Abelian fluxes, namely,
\begin{equation}
\rho_2 > 2 \lambda (\lambda^2 - f^2|\eta|^2 - 3J^2) > 0.
\end{equation}
Let us now choose a solution to the above equations as
$\lambda^2 = f^2|\eta|^2 $, which implies 
\begin{equation}
J^2 = 2\lambda^2,
\qquad 
\rho_2 >  12|\lambda|^3 > 0.
\label{solution}\end{equation}
The inequality imposes a restriction on $|\lambda|$.
The contribution of this pair of stacks 
to the D3-brane tadpole is more than $12 |\lambda|^3$.
But the D3-brane tadpole is also bounded above by 16 by \eq{d3tadpole},  
Moreover, as we discuss shortly, the quantization
conditions require the value of $|\lambda|$ to be integral. It follows that
the maximum allowed value of $|\lambda|$ is 1 for this model.

Let us now return to the quantization conditions.
In order to keep the quantization simple but capturing
all of its non-trivial features, we choose the D9-branes
wrapping exactly once around $X$, by choosing 
$e_{\alpha} = \delta^i_{\alpha}$ in \eq{gothw}, as mentioned earlier. 
The quantization conditions restrict the fluxes through 
each of the holomorphic cycles. Denoting the unit flux quanta through the $(1,1)$- and 
$(2,2)$-cycles by $\mint$ and $\nint$, respectively,  we now write down the conditions
on the different parameters. For the $(1,1)$-cycles
associated with $\h^{+}_{ij}$ and 
$\h^{-}_{ij}$, respectively, they read
\begin{equation}
\begin{split}
2\lambda\delta_{ij} &= \mint^{+}_{ij} = 
\mint^{+}\delta_{ij},\\
2f \eta_k \epsilon_{ijk} &= \mint^{-}_{ij} = \epsilon_{ijk} 
\mint^{-}_k.
\end{split}
\end{equation}
Similarly, the conditions on the fluxes through the $(2,2)$-cycles
associated to $\H^{+}_{ij}$ and $\H^{-}_{ij}$ are
\begin{equation} 
\begin{split}
2[(f\eta_i)(f\eta_j) -\lambda^2\delta_{ij}] &= \nint^{-}_{ij}
= \epsilon_{ijk}\nint^{-}_{k}, \\
2\lambda (f\eta_k)\epsilon_{ijk} &= \nint^{+}_{ij} = 
\epsilon_{ijk}\nint^{+}_{k}, \\
g^2 \xi^k\epsilon_{ijk} &= \hat{\nint}^{+}_{ij}
= \epsilon_{ijk}\hat{\nint}^{+}_k.
\end{split}
\end{equation} 
Finally, the flux through $X$ itself
is also quantized for each of the stacks and that leads to
a pair of further constraints on the parameters, namely,
\begin{equation}
2\lambda (\lambda^2 - f^2|\eta|^2) = \rho_1, \quad
2g^2 f(\xi\cdot\eta) = \goth{\rho}_2.
\end{equation}
The quantization conditions on the fluxes imply that 
all the quantities $(\mint^{+}_{ij}, \mint^{-}_{ij} )$, 
$(\nint^{-}_{ij}, \nint^{+}_{ij}, \hat{\nint}^{+}_{ij})$ 
and $(\goth{\rho}_1, \goth{\rho}_2)$ are integers.

Let us now consider the solution $\lambda^2 = f^2|\eta|^2$. 
The quantization conditions impose the following restrictions,
\begin{gather}
2\lambda = \mint^{+},
\quad
2f\eta_k = \mint^{-}_k,
\nonumber\\
2\lambda(f\eta_k) = \nint^{+}_k, \quad
g^2\xi^k = \hat{\nint}^{+}_k ,
\quad
\rho_1=0, \quad
2g^2f(\xi\cdot\eta) = \rho_2,
\end{gather}
and,
\begin{equation} 
2\begin{pmatrix}
-\lambda^2 + (f\eta_1)^2&
(f\eta_1)(f\eta_2)&
(f\eta_1)(f\eta_3)\\
(f\eta_2)(f\eta_1)&
-\lambda^2 + (f\eta_2)^2&
(f\eta_2)(f\eta_3)\\
(f\eta_3)(f\eta_1)&
(f\eta_3)(f\eta_2)&
-\lambda^2 + (f\eta_3)^2
\end{pmatrix}
=
\begin{pmatrix}
\nint^{-}_{11}&
\nint^{-}_{12}&
\nint^{-}_{13}\\
\nint^{-}_{21}&
\nint^{-}_{22}&
\nint^{-}_{23}\\
\nint^{-}_{31}&
\nint^{-}_{32}&
\nint^{-}_{33}
\end{pmatrix}
\end{equation} 
One simple choice that satisfies all these quantization condition
is $\lambda^2 = f^2$,  $\eta_k = (1,0,0)$, and $\xi^k = (2,0,0)$. 
For this choice the quantized fluxes are, 
\begin{equation} 
\label{2cycle}
\begin{split}
\lambda = -\frac{1}{2}\mint^+,\quad
f = \frac{1}{2}\mint^{-}_1,\qquad\\
(\mint^{-}_1)^2 = (\mint^{+})^2,\quad
\mint^{-}_2 = \mint^{-}_3 = 0,
\end{split}
\end{equation} 
\begin{equation}
\label{4cycle1}
\nint^{-}_{ij} = \begin{cases}
-\frac{1}{2}(\mint^{+})^2, &\text{for}\quad i=j=2\,\text{or}\,3\\
0, &\text{otherwise.}
\end{cases}
\end{equation} 
\begin{equation} 
\label{4cycle2} 
\nint^{+}_i = 
\begin{cases}
\frac{1}{4}(\mint^{+})^2, &\text{for}\quad i=1\\
0, &\text{otherwise.}
\end{cases}
\end{equation} 
\begin{equation} 
\label{4cycle3}
\hat{\nint}^{+}_i = \begin{cases}
2g^2 &\text{for}\quad i=1,\\
0, &\text{otherwise}
\end{cases}
\end{equation} 
\begin{equation} 
4 g^2 f = \hat{\nint}^{+}_1 \mint^{-}_1 = 
		\goth{\rho}_2. 
\label{6cycle}
\end{equation} 
We also choose $\lambda=-f$, which implies $\mint^{+}=\mint_1^{-}$. 
The equation \eq{6cycle} exhibits the contribution to the
D3-brane tadpole for each such pair of stacks.

Although this model is consistent,  in order to stabilize
\emph{all} the K\"ahler moduli one needs to consider a few more of such
stacks. However, there is a rather stringent restriction on the
number of such pairs of stacks arising  from the 
D3-brane tadpole cancellation condition. 
The condition \eq{4cycle1} dictates $\mint^{+}$
to be an even integer and so its minimum value
is $\mint^{+}=2$.  From \eq{solution} and \eq{2cycle}, on the other hand, 
we derive, $\goth{\rho}_2 > (3/2)(\mint^{+})^3$,
which implies that the contribution to the D3-brane tadpole
is $\rho_2 > 12$. Since the maximum contribution to the 
D3-brane tadpole is 16 we have only one  such 
pair of stacks. There exists a solution with $\lambda^2=f^2$ and 
$\eta = (1,1,0)$, where one can introduce two such pairs of stacks
which is more suitable, though not sufficient, 
for further moduli stabilization. We would like to point out
that this restriction on the number of stacks depends on
the particular model.
In fact, in next subsection we introduce appropriate modifications
to allow a larger set of brane with fluxes to stabilize several 
complex and K\"ahler  moduli.

\subsection{Model II }  \label{modelII}
We now go on to present another model, again with two $U(2)$ stacks. 
However, this models has
constant non-Abelian fluxes in both the stacks. Let us denote 
the bundles corresponding
to the fluxes in the first and the second stack by 
$\E$ and $\hat{\E}$, respectively. The fluxes are chosen as
\begin{gather}
g_{ij} = \lambda\delta_{ij}\II_2  + \frac{\xi^a}{2}g\delta_{ij}T^a,\quad
f_{ij} = f\epsilon_{ijk}\eta_k + g\epsilon^a_{ij}T^a,\\
g_{ij} = -\lambda\delta_{ij}\II_2  + \frac{\xi^a}{2}\hat{g}\delta_{ij}T^a,\quad
f_{ij} = -f\epsilon_{ijk} \eta_k + \hat{g}\epsilon^a_{ij}T^a,
\end{gather}
which in terms of the basis $(1,1)$-cycles read
\begin{equation} 
\begin{split}
\mF(\E) = [
\lambda\h_{kk}^{+} + f\epsilon_{klm}\eta_m\h_{kl}^{-}]\II_2
+ g [\frac{\xi^a}{2}\h_{kk}^{+} + \epsilon_{akl}\h_{kl}^{-}]T^a,\\
\mF(\hat{\E}) = -[\lambda\h_{kk}^{+} + f\epsilon_{klm}\eta_m\h_{kl}^{-}]\II_2
+\hat{g}[\frac{\xi^a}{2}\h_{kk}^{+} + \epsilon_{akl}\h_{kl}^{-}]T^a
\label{model2}
\end{split}
\end{equation} 
The Chern characters of the two bundles following from these choices
are,
\begin{equation} 
\begin{split}
\label{chE}
\chz{\E} &= 2, \\
\chz{\hat{\E}} &= 2,\\
\chy{\E} &= 2[\lambda\h^{+}_{kk} + f\epsilon_{klm}\eta_m\h^{-}_{kl}],
\\ \chy{\hat{\E}} &= -\chy{\E}\\
\chx{\E} &= 2 \big[-\lambda^2\H^{-}_{kk} + f^2\eta_k\eta_l\H^{-}_{kl}
+ f\lambda\epsilon_{klm}\eta_m\H^{+}_{kl} + \frac{g^2}{2} \epsilon_{ijk}\xi^k
\H^{+}_{ij}\big],
\\
\chx{\hat{\E}} &= 2\big[-\lambda^2\H^{-}_{kk} + f^2\eta_k\eta_l\H^{-}_{kl}
+ f\lambda\epsilon_{klm}\eta_m\H^{+}_{kl} + \frac{{\hat{g}}^2}{2}
\epsilon_{ijk}\hat{\xi}^k
\H^{+}_{ij}\big],\\
\chw{\E} &= [2\lambda\big(\lambda^2-f^2|\eta|^2\big) + 2g^2 f(\xi.\eta)]\v, \\
\chw{\hat{\E}} &= -[2\lambda\big(\lambda^2 - f^2|\eta|^2\big)
+ 2{\hat{g}}^2 f (\hat{\xi}.\eta)]\v,
\end{split}
\end{equation} 
Substituting these expressions and $\J = J\h^{+}_{kk}$ \eq{jjj} we obtain
\begin{gather}
2 J ( 3\lambda^2 - f^2 |\eta|^2 ) = 2 J^3. 
\label{jjj3}
\end{gather}
Similarly, upon substitution of the same, \eq{fff} leads to the following
pair,
\begin{gather}
g^2f(\xi\cdot\eta) + \lambda(\lambda^2 - f^2|\eta|^2 - 3 J^2 ) > 0,
\label{fff3}\\
\hat{g}^2f(\hat{\xi}\cdot\eta) +\lambda(\lambda^2 - f^2|\eta|^2 - 3 J^2 ) < 0.
\label{fff4}
\end{gather}
Finally, the contribution to the D3-brane tadpole turns out to be
\begin{gather}
\hat{\rho}_2 = 2g^2f(\xi\cdot\eta) - 2\hat{g}^2f(\hat{\xi}\cdot\eta).
\label{d3tadpoleII}
\end{gather}
As in Model I, we choose the simple solution
$\lambda^2 = f^2\eta^2$. Substituting this in \eq{jjj3}
we fix the value of $J$ to
\begin{equation}
J^2 = 2\lambda^2.
\end{equation}
Substituting this value of $J$ in \eq{fff3} and \eq{fff4} we derive
\begin{equation} 
\begin{split}
g^2f(\xi\cdot\eta) > 6 \lambda^3, \\
\hat{g}^2f(\hat{\xi}\cdot\eta) < 6\lambda^3.
\label{d3tadpole5}
\end{split}
\end{equation} 
The quantization conditions restrict each of the three quantities
$g^2f(\xi\cdot\eta)$, $\hat{g}^2f(\hat{\xi}\cdot\eta)$
and $6\lambda^3$ to be integral. Let us now consider the quantization conditions in some
detail. The fluxes through the $(1,1)$-cycles are quantized as
\begin{gather}
2\lambda\delta_{ij} = \mint^{+}_{ij} = \mint^{+}\delta_{ij},\notag\\
2f \eta_k  = \mint^{-}_k, \notag\\
2[(f\eta_i)(f\eta_j)-\lambda^2\delta_{ij}] = \nint^{-}_{ij}, \notag\\
2\lambda(f\eta_k) + g^2 \xi^k = \nint^{+}_{k}\\
2\lambda(f\eta_k) + \hat{g}^2\hat{\xi}^k = \hat{\nint}^{+}_k,\notag\\
2\lambda(\lambda^2 - f^2|\eta|^2) + 2g^2f(\xi\cdot\eta) = \rho_1\notag\\
2\lambda (\lambda^2 - f^2|\eta|^2) + 2\hat{g}^2f(\hat{\xi}\cdot\eta) =
-\hat{\rho}_1.\notag
\end{gather}
These conditions lead to fixing the different parameters to discrete values as,
\begin{gather}
\lambda = \frac{1}{2}\mint^{+}, \notag\\
f\eta_i = \frac{1}{2}\mint^{-}_i, \nonumber\\
g^2\xi^k = \nint^{+}_k - \frac{1}{2}\mint^{+}\mint^{-}_k,\notag\\
{\hat{g}^2}\xi^k = \hat{\nint}^{+}_k -\frac{1}{2}\mint^{+}\mint^{-}_k,\\
\nint^{-}_{ij} = \frac{1}{2}[\mint^{-}_i\mint^{-}_j-(\mint^{+})^2\delta_{ij}]\nonumber\\
\rho_1= \frac{1}{4}\mint^{+}[(\mint^{+})^2-(\mint^{-})^2] +\mint^{-}_k[
\nint^{+}_k-\frac{1}{2}\mint^{+}\mint^{-}_k] > 0,\nonumber \\
-\hat{\rho}_1= \frac{1}{4}\mint^{+}[(\mint^{+})^2 - (\mint^{-})^2]+\mint^{-}_k[
\hat{\nint}^{+}_k - \frac{1}{2}\mint^{+}\mint^{-}_k] < 0 . \notag
\end{gather}
We have already chosen a  solution, namely, 
$\lambda^2 = f^2\eta^2$. Let us also choose $\eta_k = (1,0,0)$ 
and $\xi^k =\hat{\xi}^k = (2,0,0)$.
We then have $\mint^{+} = \mint^{-}_1$, 
$\mint^{-}_2 = \mint^{-}_3=0$,
$2g^2=\nint^{+}_1 - \frac{1}{2}(\mint^{+})^2$
and
$2\hat{g}^2=\hat{\nint}^{+}_1 - \frac{1}{2}(\mint^{+})^2$.
Since each non-zero entry of the matrix $\nint^{-}_{ij}$ 
is an integer, $\mint^{+}=\mint^{-}_1$ has 
to be an even integer too. 
Therefore it follows from the above equations 
that both $g^2 \xi^k$ and $\hat{g}^2 \hat{\xi}^k$
are integers. Further,
substituting the values of $\lambda$, $f$, $\xi$, 
$\hat{\xi}$ and $\eta$ in \eq{d3tadpole5} we obtain,
\begin{gather}
\rho_1 = 2\mint^{+} g^2 > 
\frac{3}{2}(\mint^{+})^3, \qquad
\hat{\rho}_1 = -2\mint^{+} \hat{g}^2 
> - \frac{3}{2}(\mint^{+})^3,
\label{rho101}\end{gather}
so that the minimal choice is
\begin{gather}
\rho_1 = \frac{3}{2}(\mint^{+})^3  + 2, \qquad
\hat{\rho}_1= -\frac{3}{2}(\mint^{+})^3  + 2.
\label{rho102}\end{gather}
Thus, the contribution to the D3-brane tadpole is
\begin{equation}
\goth{\rho}_2 = \goth{\rho}_1 + \hat{\goth{\rho}}_1 =4.
\label{rho2}\end{equation}
Now from \eq{rho101} and \eq{rho102} we can write down the value
of $g^2$ and ${\hat{g}}^2$ as, 
\begin{gather}
2g^2 = \frac{3}{4}(\mint^{+})^3 + \frac{2}{\mint^{+}},\quad
2{\hat{g}}^2 = \frac{3}{4}(\mint^{+})^3 - \frac{2}{\mint^{+}}.
\end{gather}
Since $2g^2$ and $2{\hat{g}}^2$ are integers, and $\mint^{+}$ is an 
even integer, the only possible solution is
\begin{gather}
\mint^{+}=2,\quad g^2 = \frac{7}{2},\quad\hat{g}^2 = \frac{5}{2}.
\end{gather}

As the maximum allowed value of D3-brane tadpole is 16
from \eq{rho2} we conclude that this model can accommodate 4 
such pairs of stacks.

We now demonstrate that three such pairs of stacks
can stabilize \emph{all} the complex structure moduli.
Let us consider three pairs of stacks with fluxes such that
$\lambda^2=f^2$ for each and
\begin{equation} 
\begin{split}
\xi^{(1)} &= \hat{\xi}^{(1)}= 2\eta^{(1)} = (2,0,0),\\
\xi^{(2)} &= \hat{\xi}^{(2)}=2\eta^{(2)} = (0,2,0),\\
\xi^{(3)} &= \hat{\xi}^{(3)}=2\eta^{(3)} = (0,0,2),
\label{3stacks}
\end{split}
\end{equation} 
where $\eta^{(a)}$ denotes the value of the parameter $\eta$ in the $a$-th pair.
Substituting these in the expression of ${\mathscr F}^{(2,0)}$
given in \eq{F20} and demanding that it vanishes, it turns out
that the Abelian part is restrictive enough to fix all the
complex structure moduli and so we consider only an Abelian
flux. Thus we  obtain,
\begin{equation} 
f\epsilon_{klm}\eta_m {\bar\I}_{ki}{\bar\I}_{lj} + f\eta_m\epsilon_{ijm} 
= \lambda({\bar\I}_{ji} - {\bar\I}_{ij}).
\label{F201}
\end{equation}
Since ${\bar\I}$ is an invertible matrix, we can rewrite the first term in this
equation as,
\begin{equation} 
\begin{split}
\epsilon_{klm}{\bar\I}_{ki}{\bar\I}_{lj}i
&=\epsilon_{kln}{\bar\I}_{ki}{\bar\I}_{lj}
({\bar\I}_{np}{\bar\I}^{-1}_{pm}), \\
&=\triangle\,{\bar\I}\epsilon_{ijp}{\bar\I}^{-1}_{pm},
\end{split}
\end{equation} 
where $\triangle=\det{\bar\I}$ to simplify \eq{F201} to, 
\begin{equation} 
f\epsilon_{iln}[\delta_{nm} + \triangle{\bar\I}^{-1}_{nm}]\eta_m 
= \lambda({\bar\I}_{li} - {\bar\I}_{il}).
\label{F202} 
\end{equation} 
Corresponding to the three choices of
$\eta$, that is,  $\eta^{(1)}$, $\eta^{(2)}$ and $\eta^{(3)}$ in \eq{3stacks},
equation \eq{F202} gives rise to three equations, namely,
\begin{equation} 
\begin{split}
\delta_{n1} + \triangle {\bar\I}^{-1}_{n1} 
&= \frac{1}{2} \epsilon_{iln}({\bar\I}_{li} - {\bar\I}_{il}), \\
\delta_{n2} + \triangle {\bar\I}^{-1}_{n2}  
&= \frac{1}{2} \epsilon_{iln}({\bar\I}_{li} - {\bar\I}_{il}),\\
\delta_{n3} + \triangle {\bar\I}^{-1}_{n3}  
&= \frac{1}{2} \epsilon_{iln}({\bar\I}_{li} - {\bar\I}_{il}).
\end{split}
\end{equation} 
Using these equations we construct the $3\times 3$ matrix
${\bar\I}^{-1}$ as
\begin{equation} 
{\bar\I}^{-1} =
\begin{pmatrix}
a - \triangle^{-1}&a&a\\
b&b-\triangle^{-1}&b\\
c&c&c-\triangle^{-1}
\end{pmatrix},
\label{F203}
\end{equation} 
where $a$, $b$, and $c$ are defined as,
\begin{gather} 
({\bar\I}_{32} - {\bar\I}_{23}) = a\triangle,
\quad
({\bar\I}_{13} - {\bar\I}_{31}) = b\triangle,
\quad
({\bar\I}_{21} - {\bar\I}_{12}) = c\triangle.
\label{F207}
\end{gather}
Equating the determinant of the matrix
on the right side of \eq{F203} to
$\triangle^{-1}$
we derive a relation among $a$, $b$, $c$, as
\begin{equation}
a+b+c=\triangle + \triangle^{-1}.
\label{F205}
\end{equation}
Now we invert the matrix in (\ref{F203}) to obtain
the conjugate of the complex structure matrix,
\begin{equation} 
{\bar\I} = 
\begin{pmatrix}
a-\triangle&a&a\\
b&b-\triangle&b\\
c&c&c-\triangle
\end{pmatrix},
\label{F206}
\end{equation} 
where we have the relation used \eq{F205}.
Upon substituting the entries of the matrix
$\bar\I$ from \eq{F206} into 
\eq{F207}, self-consistency of these expressions leads to three equations as,
\begin{equation} 
\begin{pmatrix}
1&-1&\triangle\\
\triangle&1&-1\\
-1&\triangle&1
\end{pmatrix}
\begin{pmatrix}
a\\b\\c
\end{pmatrix} =0.
\label{F208}
\end{equation} 
Summing the three equations we derive
$a+b+c=0$ and therefore, by \eq{F205}, we get
\begin{equation} 
\triangle + \triangle^{-1} = 0.
\label{F209}
\end{equation} 
Thus, we have two possible values of 
$\triangle$, \viz $\triangle = \pm i$.
Noting that the  only solution to the equation \eq{F208} is
$a=b=c=0$, we derive from
\eq{F206} that $\I_{ij}=  \mp i\delta_{ij}$. Thus, 
all the complex structure moduli are stabilized upto a 
sign in this model. 
Between these, only  $\I= i\II_3$ is physically acceptable.

Let us now examine how many of the K\"ahler moduli can be
stabilized in this model.
For this, let us turn on fluxes with the parameters chosen above 
in the two stacks $\E$ and $\hat{\E}$. Substituting  \eq{model2} 
with these parameters
in \eq{jjj}, we find two equations  
from the stacks $\E$ and $\hat{\E}$, respectively. 
\begin{gather}
\lambda^2\J_{i\bar\imath} - f^2(\J_{i\bar\jmath}\eta_i\eta_j)
 = \det\J +
(\lambda f\eta_k + g^2 \xi^k)\epsilon_{ijk}\J_{i\bar\jmath} = 0,
\label{stableI}\\
\lambda^2\J_{i\bar\imath} - f^2 (\J_{i\bar\jmath}\eta_i\eta_j)
 = \det\J +
(\lambda f\eta_k + \hat{g}^2 \hat{\xi}^k)\epsilon_{ijk}\J_{i\bar\jmath} = 0,
\label{stableII}
\end{gather}
to be simultaneously satisfied to fix the K\"ahler form $\J$.
Subtracting one from the other of these equations we obtain,
\begin{equation}
(g^2\xi^k - {\hat{g}}^2{\hat\xi}^k)\epsilon_{ijk}\J_{i\bar\jmath}=0.
\label{stable3}\end{equation}
We now substitute the values of the parameters, determined above,
$g^2=7/2$, $\hat{g}^2=5/2$ and the following three 
values of $\xi$
corresponding to the three pairs of stacks introduced in \eq{3stacks},
\begin{equation}
\xi^{(1)} = (2,0,0),\quad\xi^{(2)} = (0,2,0),\quad
\xi^{(3)} = (0,0,2),
\end{equation} 
in equation \eq{stable3} and derive,
$\J_{1\bar2}=\J_{2\bar1}$, $\J_{2\bar3}=\J_{3\bar2}$, $\J_{3\bar1}=\J_{1\bar3}$.
Thus, $\J$ is fixed to be a symmetric matrix.
Upon setting the antisymmetric part of $\J$ equal to zero 
reduce \eq{stableI} and \eq{stableII}
into a single equation,
\begin{equation}
\lambda^2\J_{i\bar\imath} - f^2\,\J_{i\bar\jmath}\eta_i\eta_j
 = \det\J.
\label{stable4}\end{equation}
Finally, substituting $\lambda^2 = f^2|\eta|^2$ in \eq{stable4}, 
along with the following three 
values of $\eta$
corresponding to the three pairs of stacks \eq{3stacks},
\begin{equation} 
\eta^{(1)} = (1,0,0),\quad\eta^{(2)} = (0,1,0),\quad\eta^{(3)} = (0,0,1),
\end{equation} 
we get equations:
\begin{equation} 
\lambda^2 (\J_{2\bar2} + \J_{3\bar3}) =
\lambda^2 (\J_{3\bar3} + \J_{1\bar1}) =
\lambda^2 (\J_{1\bar1} + \J_{2\bar2}) = \det\J,
\end{equation} 
which implies, in turn, that
\begin{equation} 
\label{diageq}
\J_{1\bar1} = \J_{2\bar2} = \J_{3\bar3} \equiv J.
\end{equation} 
This exhausts all the relations available for fixing the moduli.
Thus, in this model three parameters from 
the symmetric part of $\J$ are not 
fixed. 

Let us indicate a possible set up for overhauling this stabilization scheme by
fixing the symmetric part of $\J$ too.
However, let us point out that 
tadpole bounds can not be met in trying to fix the K\"ahler moduli in totality.
We can stabilize the off-diagonal terms of the symmetric matrix $\J$
to zero by introducing three further stacks of branes.
Taking the liberty of introducing
three more stacks with $\eta$ given by,
\begin{equation} 
\eta^{(4)} = (1,1,0),\quad\eta^{(5)} = (1,0,1),\quad\eta^{(6)} = (0,1,1),
\end{equation} 
perhaps at the risk of violating the the tadpole conditions, and substituting these 
in \eq{jjj} we derive,
\begin{equation} 
\lambda^2 (2J - \J_{1\bar2}) =
\lambda^2 (2J - \J_{2\bar3}) =
\lambda^2 (2J - \J_{3\bar1}) = \det\J,
\end{equation} 
where we have used equation \eq{diageq} and the fact that $\J$ is 
symmetric.
These equations set all the off-diagonal terms
to zero upon using $\det\J = 2\lambda^2 J$.
Thus, these three additional pairs of stacks
stabilize all the K\"ahler moduli
$\J=iJ_{i\bar\jmath}dz^i\wedge d{\bar z}^{\bar\jmath}$ 
to $\J = J\h_{kk}^{+}$.
\section{Discussion}\label{secdis}
In this article we have discussed a scheme for moduli stabilization
in an orientifold  $\T^6/\Omega\cdot\Z_2\cdot(-1)^{F_L}$ of
the type--IIB theory
using D9-branes with a non-Abelian magnetic flux on the
world-volume of the brane. Unlike its Abelian counterpart, derivation
of the supersymmetry condition, though 
possible\cite{Marino:1999af}, has not been done explicitly 
from open string theory. We use the supersymmetry condition
proposed in the context of BPS branes in an $\N=2$ theory. 
We use this supersymmetry preserving condition in the
toroidal orientifold. We have presented two simple models
and demonstrated that properly chosen non-Abelian 
fluxes stabilize most of the moduli.
As is well known, there is a restriction
ensuing from D3-brane tadpole cancellation
condition which limits the number of allowed
D-brane severely. Therefore, although in the second
model all the complex structure moduli are stabilized,
the only some of the K\"ahler moduli could be stabilized in compliance with the tadpole
constraints, a complete stabilization being prevented by the dearth of branes allowed.

In both the models that we analyze we obtain the value of
the K\"ahler volume to be of the order of the string scale.
In these  models the arbitrarily small
values of the K\"ahler volume can be attained by scaling the 
wrapping numbers. Though we have not studied it, a natural candidate to
obtain large K\"ahler volume seems to be the T-dual of the present
version, namely the type--I string theory on the orientifold 
\cite{Antoniadis:2004pp,Blumenhagen:2005zh}.

We have not considered the issues regarding the stability of 
the bundles used for stabilization. As the volume turns out to
be small it will be prudent to check the $\Pi$-stability of the
bundles which is valid all over the K\"ahler moduli
space. However, in order to check $\Pi$-stability one needs 
exact expressions of the central charge, which
is not available in the present case at the moment. For some of the 
Calabi-Yau manifolds, though, one can write down 
exact expressions of central charges associated with
the D-brane configuration from the periods of the
mirror manifolds. Thus, Calabi-Yau manifolds
are interesting candidates for generalizations of this scheme 
of moduli stabilization.
As we have explored here only 
a fraction of the different possibilities, it may be
possible, even within the current set up, to find stable bundles
that stabilize the K\"ahler volume to a large value.

Though we showed the stabilization of complex structure 
moduli, the axion-dilaton modulus remains unfixed. 
In addition to the present configuration, 
one can \cite{Kachru:2002he} turn on NS-NS and RR
fluxes through three-cycles and thus stabilize the axion-dilaton
modulus as has been done in the context of Abelian fluxes.
However, turning on three-form fluxes in the presence of
brane may involve Freed-Witten anomaly \cite{Freed:1999vc} which 
need to be taken care of. For an analysis of Supersymmetric 
D-branes in presence of background RR fluxes see \cite{Martucci:2005ht}.

A natural extension
of this work is to study supersymetry breaking in this 
formulation. A mechanism of supersymmetry breaking by generating 
D-terms has already been discussed\cite{Antoniadis:1997mm,Villadoro:2006ia}
for D-branes with Abelian magnetic fluxes. 
It is interesting
to study this (and other) supersymmetry breaking mechanism 
to the case of non-Abelian fluxes. A statistical
measure of the realistic vacua for this class of moduli 
stabilization along the line of\cite{Ashok:2003gk,Gmeiner:2005vz},
as well as a study of open string moduli\cite{Gomis:2005wc} 
will also be of 
relevance for a better understanding of such string backgrounds.
\appendix
\section{Appendix}\label{appndx}
In this Appendix we lay down the conventions and notations we used in the main text.
Complex co-ordinates of $X$, namely,  $z^i$, ${\bar z}^{\bar\imath}$ are related 
to the real co-ordinates $x^i$, $y^i$ through
$z^i = (x^i + \I_{ij}y^j)$, $i, j =1,2,3$.
The $3\times 3$ matrix $\I_{ij}$ with complex entries
stands for the complex structure of $X$.
We use unbarred indices such as $i,j, \cdots$ for the  holomorphic co-ordinates
and barred indices such as ${\bar\imath}, {\bar\jmath}, \cdots$ 
for the antiholomorphic ones.

In the complex basis of co-ordinates, the two-form field 
strengths are denoted $\mF^{(2,0)} = \mF_{ij} dz^i\wedge dz^i $,
$\mF^{(1,1)} = \mF_{i\bar\jmath} dz^i\wedge d\bar{z}^{\bar\jmath} $, etc.,
while in the basis of real forms, they are denoted $F_{x^ix^j}$.
The K\"ahler two-form of the compactification manifold
$X$ is written as 
$\J = \J_{i{\bar\jmath}}dz^i\wedge d{\bar z}^{\bar\jmath} $
We use the following notation for the basis of
$(1,1)$- , $(2,2)$- and $(3,3)$-forms, respectively,
in space-time co-ordinates,
\begin{gather}
\h_{ij}=\frac{1}{2}dz^i\wedge d{\bar z}^{\bar\jmath}, \\
\H_{ij}= - \frac{1}{4}(\frac{1}{2}\epsilon_{ikm} dz^k\wedge dz^m)
\wedge
(\frac{1}{2}\epsilon_{jln} d{\bar z}^{\bar l}\wedge d{\bar z}^{\bar n}), \\
\v = 
-\frac{i}{8}\,\Pi_{i=1}^3 (dz^i\wedge d{\bar z}^{\bar\jmath})
\end{gather}
The elements of the above basis for $(1,1)$-forms
and $(2,2)$-forms are complex. 
We introduced another basis in which the even dimensional 
forms are also hermitian,
\begin{gather}
\h^{-}_{ij} = \frac{1}{2}[\h_{ij} - \h_{ji}],\\
\h^{+}_{ij} = \frac{i}{2}[\h_{ij} + \h_{ji}],\\
\H^{-}_{ij} = \frac{1}{2}[\H_{ij} + \H_{ji}], \\
\H^{+}_{ij} = \frac{i}{2}[\H_{ij} - \H_{ji}].
\end{gather}
The $(1,1)$- and $(2,2)$-forms are related as,
\begin{gather} 
\h_{ij}^{-} \wedge \h_{kl}^{-}
= \frac{1}{2}(\epsilon_{ikm}\epsilon_{jln}
- \epsilon_{ilm}\epsilon_{jkn})\H_{mn}^{-},\\
\h_{ij}^{-}\wedge\h_{kl}^{+}
= \frac{1}{2}(\epsilon_{ikm}\epsilon_{jln}
+ \epsilon_{ilm}\epsilon_{jkn})\H_{mn}^{+},\\
 \h_{ij}^{+} \wedge \h_{kl}^{+}
= -\frac{1}{2}(\epsilon_{ikm}\epsilon_{jln}
+ \epsilon_{ilm}\epsilon_{jkn})\H_{mn}^{-}.
\end{gather} 
The intersections of the these forms are, 
\begin{gather}
\h_{ij}^{-} \wedge \H_{kl}^{-} = 0
= \h_{ij}^{-} \wedge \H_{kl}^{+},\\
\h_{ij}^{+} \wedge \H_{kl}^{-} =-\frac{1}{2}(\delta_{ik}\delta_{jl} +
\delta_{il}\delta_{jk} )\v,\\
\h_{ij}^{-} \wedge \H_{kl}^{+} =-\frac{1}{2}( \delta_{ik}\delta_{jl} -
\delta_{il}\delta_{jk} )\v.
\end{gather}


\end{document}